\documentclass[mypaper,7pt,twoside]{CoAst}
\usepackage{epsf,graphicx,fancyhdr}
\renewcommand{\chaptermark}[1]{\markboth{#1}{}}

\newcommand{\kopf}{\small\itshape Comm. in Asteroseismology\\ Vol. 143, 2003}
\newcommand{\Authors}[1]{\begin{center}\normalsize\bf\sf #1 \end{center}}

\renewcommand{\author}[1]{\begin{center}\normalsize\bf\sf #1 \end{center}}
\newcommand{\Address}[1]{\begin{center}\small\sf #1 \end{center}}

\renewenvironment{abstract}{\section*{Abstract}\normalsize\sf}{}
\newcommand{\References}[1]{\begin{flushleft}{\large References\\}\vspace*{2mm}\small #1 \end{flushleft}}

\newcommand{\chapterDSSN}[2]{\chapter[\sf\normalsize #1\\ \footnotesize \hspace*{5mm}by #2 \sf\normalsize][]{#1\\}\rhead[\fancyplain{}{\sf\footnotesize \center{#1}}]{\fancyplain{}{\sffamily\thepage}}\lhead[\fancyplain{\kopf}{\sffamily\thepage}]{\fancyplain{\kopf}{\sf\footnotesize \center{#2}}}}

\newcommand{\figureDSSN}[5]{\begin{figure}[#4]
\centering
\includegraphics*[#5]{#1}
\caption{#2}
\label{#3}
\end{figure}}

\renewcommand{\chaptername}{}
\newcommand{\acknowledgments}[1]{\vspace*{5mm}\noindent\begin{bf}Acknowledgments. \end{bf} #1}

\begin{document}
\sf

\chapterDSSN{Photometry of Be stars in the vicinity of COROT primary targets for asteroseismology}{J. Guti\'errez-Soto et al.}

\Authors{J. Guti\'errez-Soto$^1$, J. Fabregat$^1$, J. Suso$^2$, A.M. Hubert$^3$, M. Floquet$^3$ \\and R. Garrido$^4$} 
\Address{$^1$ Observatori Astron\`omic, Universitat de Val\`encia\\
$^2$ Instituto Ciencia de los Materiales, Universitat de Val\`encia\\
  $^3$GEPI, Observatoire de Paris-Meudon \\
$^4$Instituto de Astrof\'\i sica de Andaluc\'\i a}

\noindent
\begin{abstract}

We present differential photometry of Be stars close to potential COROT primary targets for asteroseismology. Several stars are 
found to be short period variables. We propose them to be considered as secondary targets in the COROT asteroseismology fields.
\end{abstract}

\section{Introduction}
The observation of classical Be stars by COROT will provide important keys to understand the physics of  these objects and the nature of  the 
Be phenomenon.In particular, the detection of photospheric multiperiodicity will confirm the presence of non radial pulsations ($nrp$) as the 
origin of the short term variability. COROT observations will allow the study of the beat phenomenon of $nrp$ modes and its 
relation with recurrent outbursts and the building of the circumstellar disc. 

Our group is proposing the observation of Be stars as secondary targets for the asteroseismology fields. 
A sample of stars in the vicinity of the main target candidates is under study for this purpose. Hubert et al. (2001, 2003) 
presented the selected objects and performed a study of their short term variability using Hipparcos photometric data. 
We have obtained new ground based photometry with a more suitable time sampling to further characterize their variability.

\section{Observations and data analysis}

Observations were done at the 0.9 m telescope of the Observatorio de Sierra Nevada (Granada, Spain). 
The instrument used was the automatic six-channel spectrophotometer which allows simultaneous observations 
through the four $uvby$ filters of the Str\"omgren system. Differential photometry of four Be stars in the 
vicinity of main targets candidates in the galactic center direction was obtained in the period 20 to 29 May 2002. 
Five more Be stars were observed in the anticenter direction in the period 8 to 14 January 2003. 
The mean accuracy of the differential photometry, measured as the rms of the comparison minus check values, is 0.010 mag. in $u$ and 0.007 in $vby$. 

Different observing strategies were employed in both runs. In May 2002 we observed all four stars consecutively every night. 
In January 2003 we devoted each clear night to follow an individual star. For the first run data we performed periodogram analysis 
of each star, using the Scargle (1982) and PDM (Stellingwerf 1978) techniques. Phase curves at the detected frequencies were made 
and visually inspected. In Figure 1 we present an example of the obtained periodograms. With the second run data, we made daily 
light curves which were visually inspected for consistent trends. Light curves of three stars are presented in Figure 2.

\figureDSSN{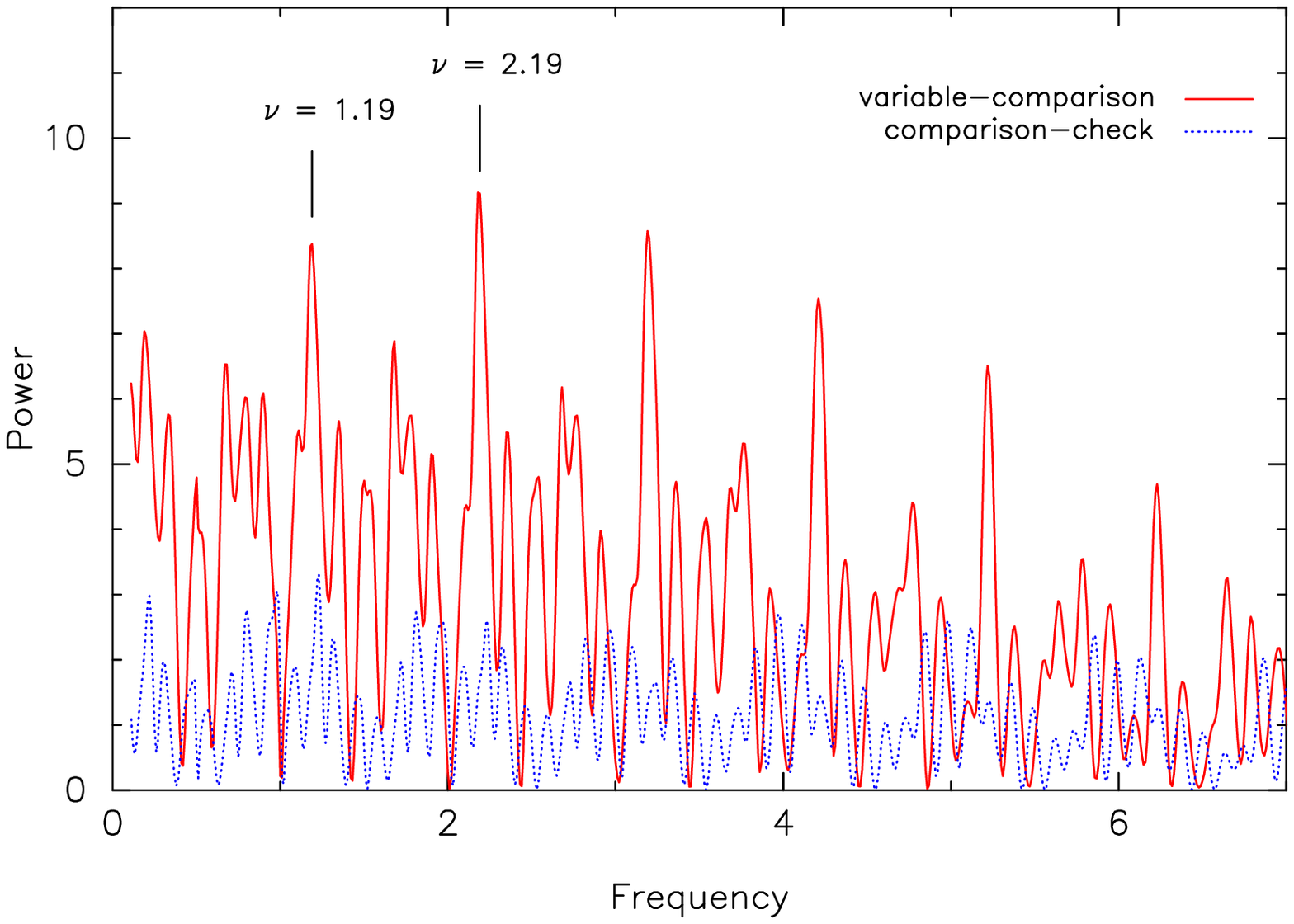}{Scargle periodogram for the v filter data of star HD 183656}{label}{!ht}{clip,angle=0,width=10cm}
\figureDSSN{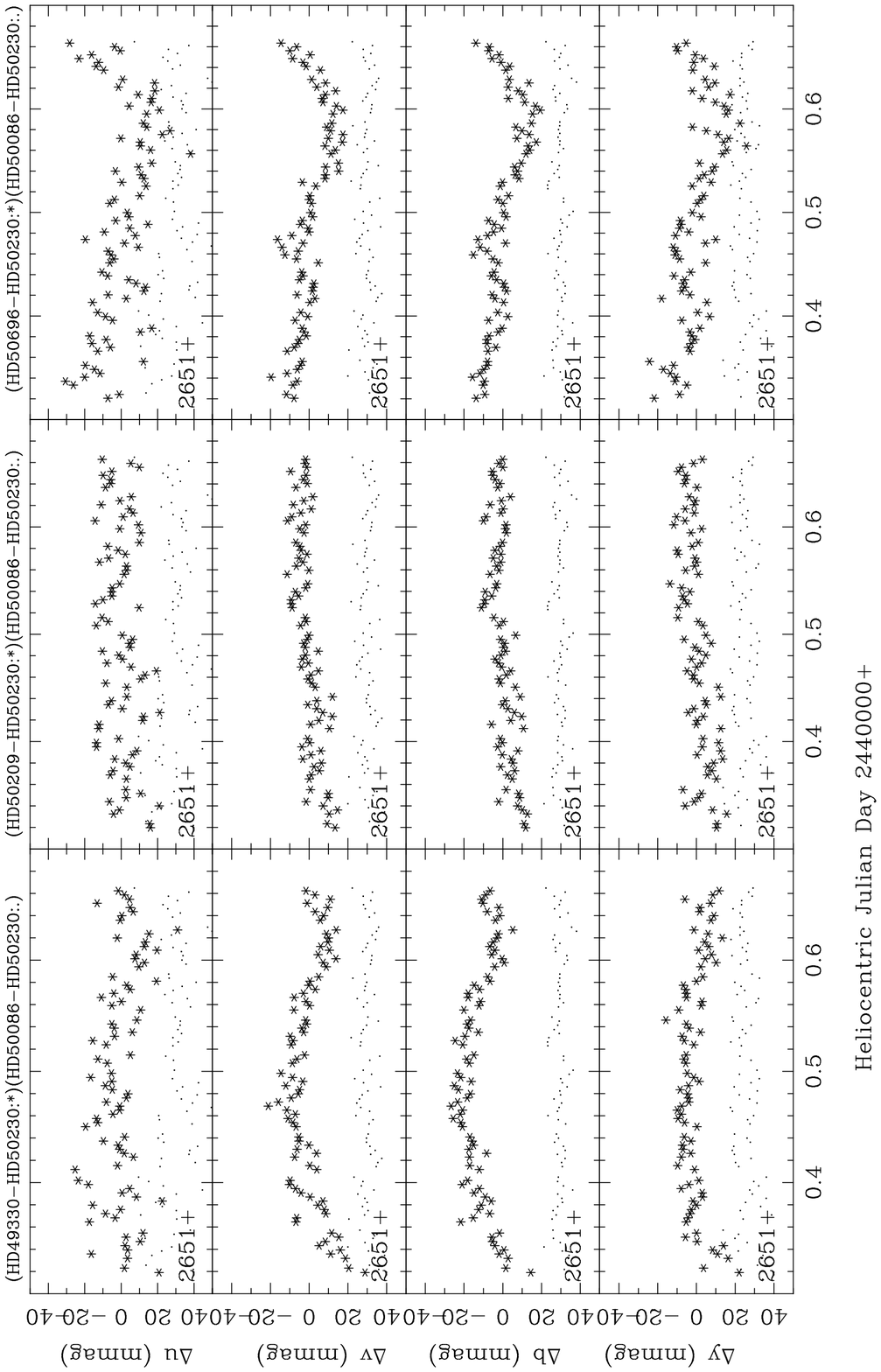}{Light curves of stars HD 49330, HD 50209 and HD 50696. Asterisks correspond to variable minus comparison magnitude, 
and points to comparison minus check.}{label}{!ht}{clip,angle=270,width=11cm}

\section{Notes on individual stars}

{\em HD 43285.} A possible period of 0.454 days is proposed from the Hipparcos data. Our light curve spanning 8.5 hours does not show any variability. \\

{\em  HD 46380.} This star is presented as short term variable from Hipparcos, although no period was found. 
Our data confirm this star as variable. Our time coverage prevents us from performing period searches. \\

{\em  HD 49330.} The Hipparcos data show a periodicity of 0.283 days. Our 8.5 hours light curve presents clear 
variability (see Figure 2). The short time coverage does not allow us to perform period searches. However, we 
have made a phase curve with the Hipparcos period, which we present in Figure 3. Our data are fully compatible with the 
Hipparcos period determination. \\

\figureDSSN{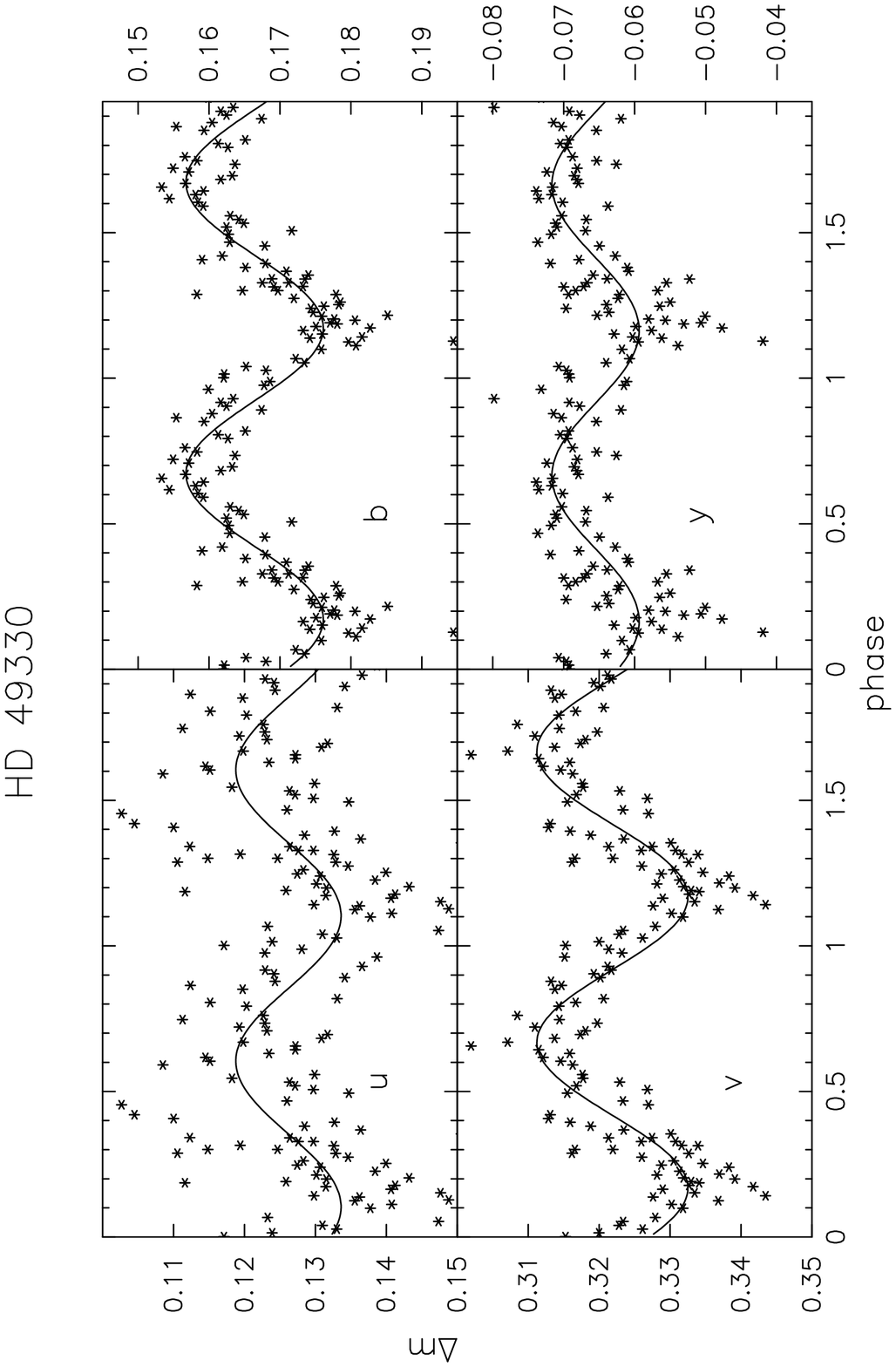}{Phase curve of star HD 49330 with period 0.283 d.}{label}{!ht}{clip,angle=270,width=11cm}
\figureDSSN{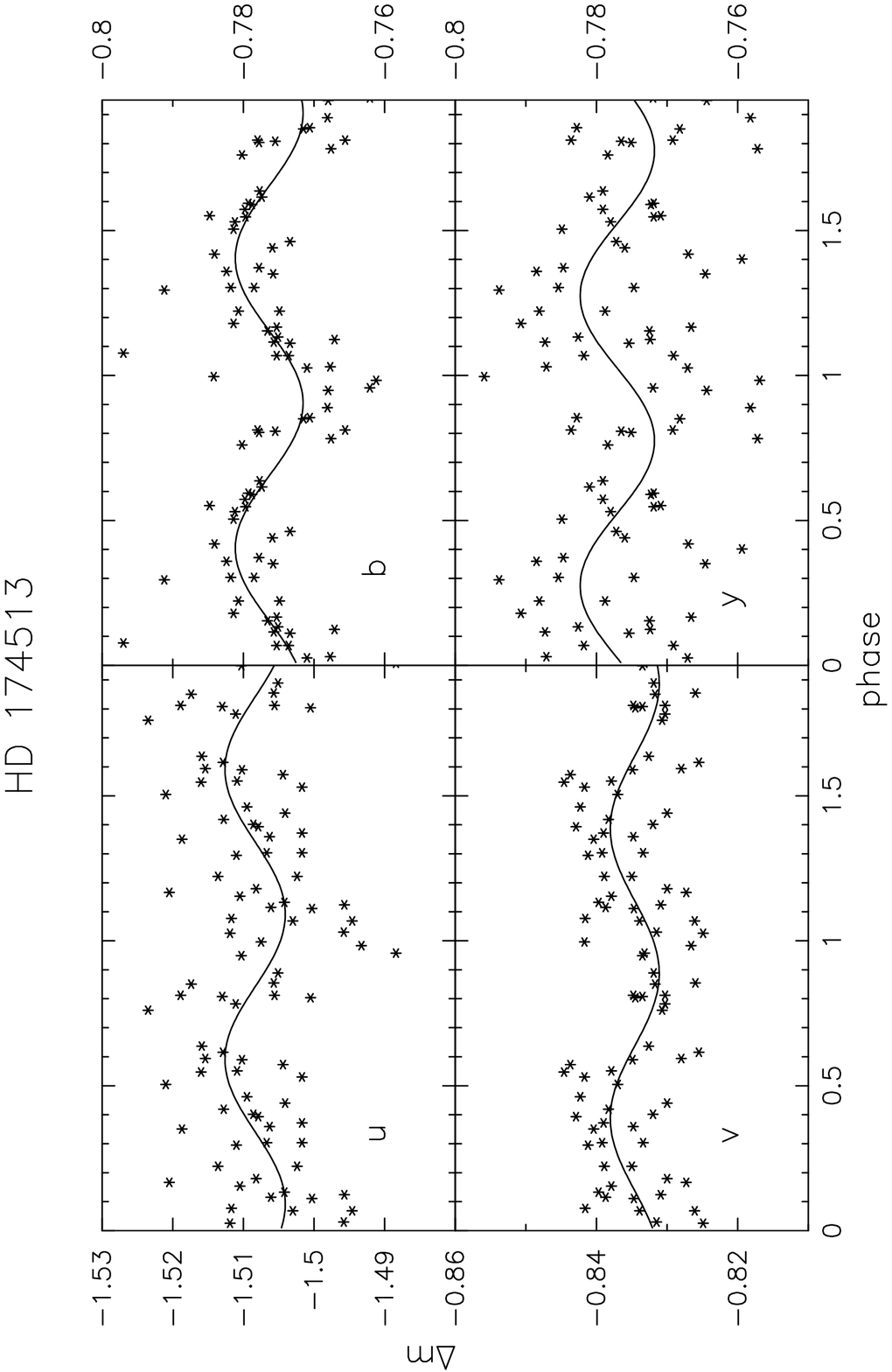}{Phase curve of star HD 174513 with period 0.234 d.}{label}{!ht}{clip,angle=270,width=11cm}

\figureDSSN{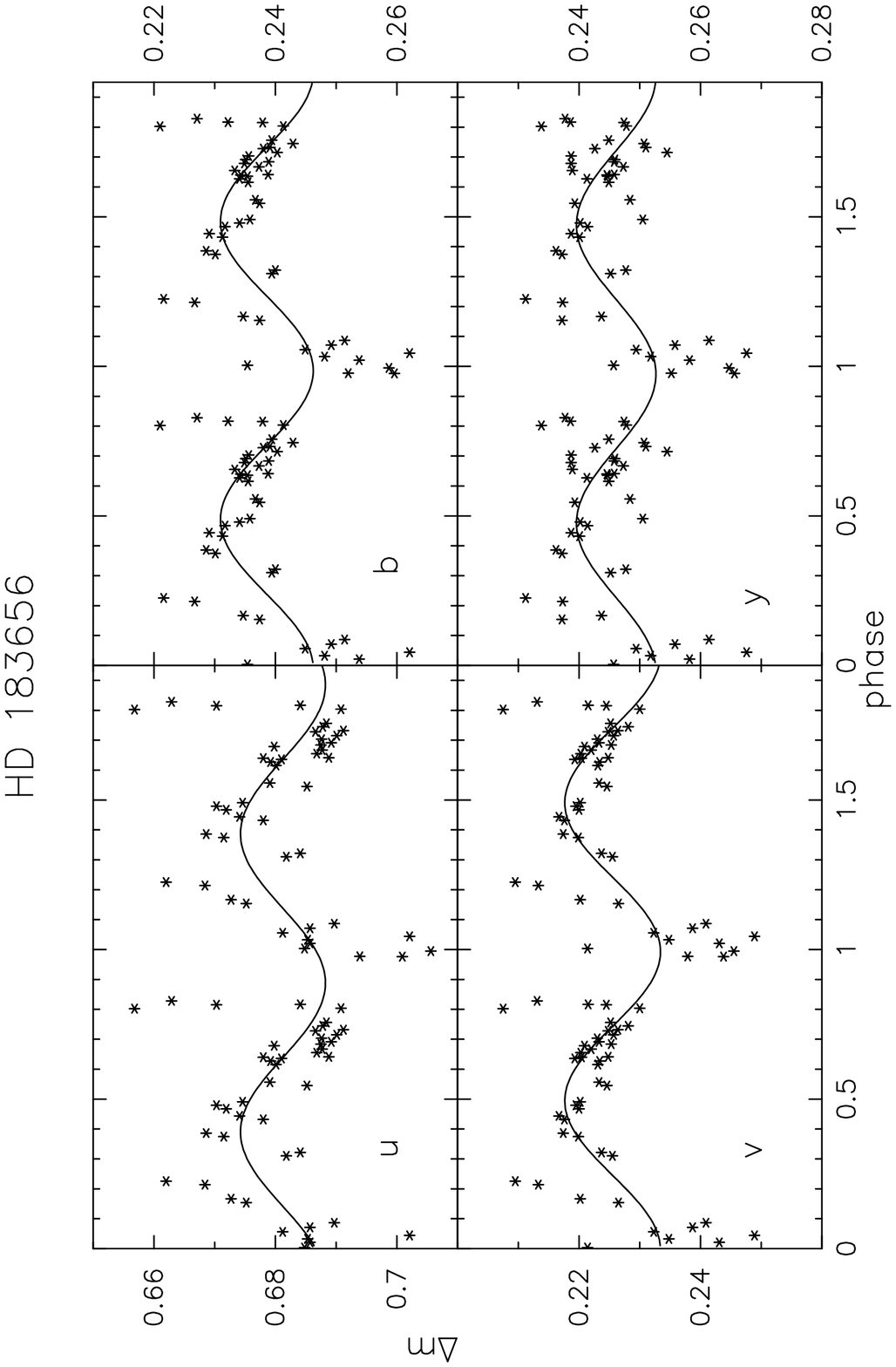}{Phase curve of star HD 183656 with period 0.8518 d.}{label}{!ht}{clip,angle=270,width=11cm}

\figureDSSN{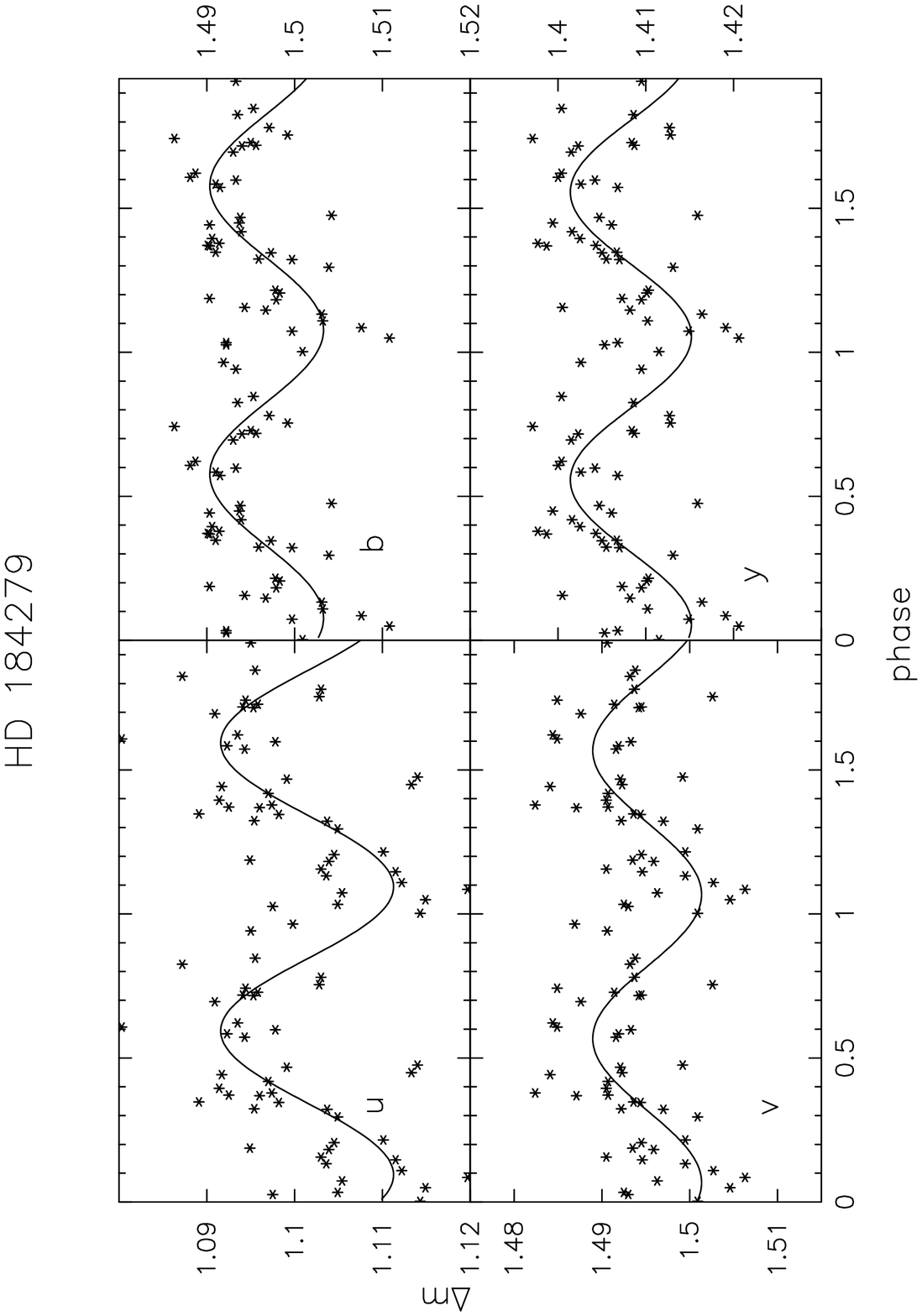}{Phase curve of star HD 184279 with period 0.423 d.}{label}{!ht}{clip,angle=270,width=11cm}

{\em HD 50209.} The Hipparcos data analysis shows a period of 0.592 days, considered as uncertain. Our data show clear variability (see Figure 2). 
The phase curve is compatible with the Hipparcos period, although the phase coverage is incomplete.\\

{\em HD 50696.} There are no Hipparcos photometric data for this faint star. Our observations show clear variability (see Figure 2).\\

{\em HD 171219.} No short term variability was found in the Hipparcos photometry. Our periodogram analysis does not show any significant peak. 
The star is not variable at our detection level.\\

{\em HD 174513.} This star appears as short term variable in the Hipparcos data, although no periodicity was found. 
Our analysis confirms the variability with possible periods of 0.234 or 0.306 days. In Figure 4 we present 
phase curves with the 0.234 days period,  which we consider the most likely one.\\

{\em HD 183656.} Short term variability with a period of 0.652 days was obtained from the Hipparcos photometry. 
This star was also found variable by Lynds (1960) who proposed a period of 0.8518 days. 
Our data are more compatible with the Lynds' period (see Figure 5). 
However, the Hipparcos data appear very noisy when folded with the 0.8518 days period. 
This might indicate that the star is multiperiodic. More observations are needed to conclude on this issue.\\

{\em HD 184279.} From the analysis of the Hipparcos data, Hubert et al. (2001) proposed a period of 0.156 days, 
while Percy et al. (2002) found a period of 0.6 days. Both studies present their period determination as uncertain. 
From our data, the most likely period is 0.423 days, although this value is also uncertain. 
We present the composition with our period in Figure 6. In the Hipparcos data this period is not apparent, 
yet a period of 0.734 days appears in addition to the 0.156 days period. The 0.734 days period (f=1.346 c/d) 
is in fact an one day alias of our 0.423 days (f=2.36 c/d) period. We can hence confirm the short 
term variability and possible multiperiodicity. The determination of the actual variability characteristics would need more observational work.

\begin{table}
\caption{Summary of results}

\begin{tabular}{llll}

\hline
Star & HIPPARCOS & This Work\\
\hline
HD 43285 & Periodic P=0.454 d ? & No variable\\

HD 46380 &
Variable &
Variable \\

HD 49330&
Periodic P=0.283 d ?&
Periodic P=0.283 d\\

HD 50209&
Periodic P=0.592 d ?&
Variable \\

HD 50696&
No data&
Variable\\

HD 171219&
No short term variability&
No variable\\

HD 174513&
Variable&
Periodic P=0.234 d ?\\

HD 183656&
Periodic P=0.652 d &
Periodic P=0.8518 d \\

HD 184279&
Periodic P=0.156 or 0.6 d ?&
Periodic P=0.423 d ?\\

\hline
\end{tabular}
\end{table}

\section{Conclusions}

Our observations confirm that stars HD 49330, HD 50209, HD 174513, HD 183656 and HD 184279 are periodic variables, with the periods given in Table 1. 
Stars HD 46380 and HD 50696 are also variable, although their periods are not yet found. HD 43285 and HD 171219 appear as no variables in our data. 
All results are summarized in Table 1.

\acknowledgments{

This research was based on data obtained at the Observatorio de Sierra Nevada, which is operated by the Consejo 
Superior de Investigaciones Cient\'\i ficas through the Instituto de Astrof\'\i sica de Andaluc\'\i a. J.F. and J.S. 
acknowledges financial support from the program ESP2001-4530-PE. R.G. acknowledges financial support from the program 
ESP2001-4528-PE. The work of J.G-S. is supported by a grant from the Spanish ``Ministerio de Educaci\'on, Cultura  y Deporte''.

}

\References{

Hubert, A.M. et al. 2001, 1st COROT Week, Vienna\\
Hubert, A.M. et al. 2003, 4th COROT Week, Marseille\\
Lynds, C.R. 1960, ApJ, 131, 390L\\
Percy J. R. et al. 2002, PASP, 114, 551\\
Scargle, J.D. 1982, ApJ, 263, 835\\
Stellingwerf, R. F. 1978, ApJ, 224, 953S\\

}

\end{document}